\documentclass[12pt]{article}
\usepackage{amssymb}

\usepackage{amsmath}

\begin{document}

\begin{center}
\vspace{1cm}{\Large {\bf Brane Content of Branes' States}}

\vspace{1cm} {\bf Ruben Mkrtchyan} \footnote{ E-mail: mrl@r.am}
\vspace{1cm}

\vspace{1cm}

{\it Theoretical Physics Department,} {\it Yerevan Physics
Institute}

{\it Alikhanian Br. St.2, Yerevan 375036, Armenia}
\end{center}

\vspace{1cm}
\begin{abstract}
The problem of decomposition of unitary irreps of (super)
tensorial (i.e. extended with tensorial charges) Poincar{\'e}
algebra w.r.t. its different subalgebras is considered. This
requires calculation of little groups for different configurations
of tensor charges. Particularly, for preon states (i.e. states
with maximal supersymmetry) in different dimensions the particle
content is calculated, i.e. the spectrum of usual Poincar{\'e}
representations in the preon representation of tensorial
Poincar{\'e}. At d=4 results coincide with (and may provide
another point of view on) the Vasiliev's results in field theories
in generalized space-time. The translational subgroup of little
groups of massless particles and branes is shown to be (and
coincide with, at d=4) a subgroup of little groups of "pure
branes" algebras, i.e. tensorial Poincar{\'e} algebras without
vector generators. At 11d it is shown that, contrary to lower
dimensions, spinors are not homogeneous space of Lorentz group,
and one have to distinguish at least 7 different kinds of preons.

\end{abstract}

\renewcommand{\thefootnote}{\arabic{footnote}}
\setcounter{footnote}0 {\smallskip \pagebreak }

\section{Introduction}

The study of supersymmetric theories leads to the change of our
understanding of space-time symmetry algebras and of space-time
itself. Instead of Poincar{\'e} algebra, which is a semidirect
product of Lorentz algebra on an Abelian algebra of vectorial
generators of space-time translations, now we have additional
"translations" by tensorial charges, which are carried by branes
\cite{TowAsk}. These charges appear in the anticommutator of
supercharges, the most general among such an algebras is that of
M-theory, where anticommutator of supercharges includes all
possible tensors, namely vector, membrane and five-brane:

\begin{eqnarray}
\left\{ \bar{Q},Q\right\} &=&\Gamma ^{\mu}P_{\nu}+\Gamma
^{\mu\nu}Z_{\mu\nu}+\Gamma ^{\mu\nu\lambda\rho\sigma}Z_{\mu\nu\lambda\rho\sigma},  \label{eq1}\\
  \mu, \nu,...
&=&0,1,2,..10. \nonumber
\end{eqnarray}

Similar algebras exist in lower dimensions, below we shall
consider the minimal algebras, i.e. algebras with minimal number
of spinors $Q$. It is shown in \cite{Tow} that many results in an
M-theory can be derived directly from algebra (\ref{eq1}).
Particularly, properties of brane states of M-theory can be
studied, which is natural, since all branes are unitary irreps of
(\ref{eq1}). Let's consider the bosonic subalgebras of these
super-Poincar{\'e} algebras, e.g. for $11d$ case (\ref{eq1}) that
is Lorentz $M_{\mu\nu}$, momenta $P_{\mu}$, membrane charge
$Z_{\mu\nu}$ and 5-brane charge $Z_{\mu\nu\lambda\rho\sigma}$. We
shall call such an algebras "tensorial Poincar{\'e}" and denote
them $(M_{\mu\nu}; P_{\mu}, Z_{\mu\nu}, ...)$. Actually they are a
semidirect products of Lorentz algebra $M_{\mu\nu}$ with Abelian
algebra of generators $P_{\mu}, Z_{\mu\nu}, ...$. So bosonic
subalgebra of M-theory is $(M_{\mu\nu}; P_{\mu}, Z_{\mu\nu},
Z_{\mu\nu\lambda\rho\sigma})$, in our notations.

The natural approach to (\ref{eq1}) from the point of view of
modern field theory is to try to construct the field theories,
invariant w.r.t. such (super)-algebras, the first step of such
approach should be the construction of their unitary irreps. That
can be achieved by Wigner's method of inducing representation from
the unitary irreps of little group \cite{Wig}, \cite{BR}. Next
will be the construction of relativistic free field equations, the
space of solutions of which will give, modulo gauge invariance,
another description of unitary irreps of tensorial Poincar{\'e}.
Already at that step the generalization of space-time will be
required, because one have to introduce a tensorial coordinates
dual to tensorial charges. Such an approach was elaborated in
\cite{Man1}, \cite{Man2} for tensorial Poincar{\'e} $(M_{\mu\nu};
Z_{\mu\nu})$ in the space with two times, particularly with the
aim of study the SO(2,10) invariance hypothesis of M-theory
\cite{Bars}. The little groups of branes are calculated for some
cases in \cite{Man1}, \cite{mrl2}.  Then interaction terms have to
be constructed, which have to maintain gauge invariances - for
conventional Poincar{\'e}  case they are often determined by that
requirement. For the simplest 4d, (actually (2+2)d) case an
interaction terms are constructed in \cite{mrl} just by such
requirement. Also, the presence of tensorial charges requires the
reconsideration of spin-statistics theorem. That theorem can be
considered as a rule, assigning the definite statistics to the
irreps of Poincar{\'e} algebra. Now, for (\ref{eq1}) type algebras
(actually for their bosonic subalgebras, i.e. tensorial
Poincar{\'e}), since the classification of irreps is substantially
different from that for usual Poincar{\'e}, one has to rederive
spin-statistics theorem, the first steps in that direction were
done in \cite{mrl2}, where spin-statistics for preons \cite{Band}
is considered.

In \cite{Vas},\cite{Vas2} the $OSp(2M)$ (conformal) invariant
approach to (free) higher spin theories is developed, on the basis
of generalized space-time. It is interesting and intriguing, that
this approach leads to the same kind of space-time, as field
theory approach of \cite{Man1} \cite{Man2}. As we shall see below,
there are more precise connections between these approaches.

The new feature of tensorial Poincar{\'e} algebras is that they
have subalgebras which itself are tensorial, or sometimes usual,
Poincar{\'e} algebras. Irreps of this algebra (and corresponding
group) can be decomposed into irreps of that subgroups. That will
be the subject of study of present paper. This should help for a
(future) study of whether superstring/M-theory can be described in
this way, as some theory in space-time with coordinates dual to
all tensorial central charges. So, we shall study what irreps of
say particle Poincar{\'e}, or another tensorial Poincar{\'e} are
making up the given irrep of given tensorial Poincar{\'e}.  Hence
the title of paper: branes content of branes'. In this paper the
problem is not solved in whole generality, but, for some
interesting cases is reduced to standard problems in group theory
(harmonic analysis) and answers are given in simple cases.
Actually even in 4d there exist another subalgebras, which we
shall call "pure branes" subalgebras (see below) with respect to
which irreps can be decomposed, also.

At d=4 tensorial Poincar{\'e} includes vector - an energy-momentum
$P_\mu$, and second-rank antisymmetric tensor - membrane (domain
wall) charge $Z_{\mu\nu}$. The corresponding susy theory includes
one Majorana spinor, with susy relation:

\begin{eqnarray}
\left\{ \bar{Q},Q\right\} &=&\Gamma ^{\mu}P_{\mu}+\frac{1}{2}
\Gamma^{\mu\nu}Z_{\mu\nu},  \label{4d}\\
  \mu,\nu,...
&=&0,1,2,3. \nonumber
\end{eqnarray}
We denote the numeric value of rhs (on the subspace when it has
definite value) by $k_{\alpha}\!^{\beta}$ (for this and similar
relations in other dimensions) , and corresponding values for
$P_{\mu}$ and $Z_{\mu\nu}$ as $p_\mu(k), z_{\mu\nu}(k)$:
\begin{eqnarray}
k=\Gamma ^{\mu}p_{\mu}(k)+\frac{1}{2}\Gamma ^{\mu\nu}z_{\mu\nu}(k)
\end{eqnarray}
One natural subalgebra is usual (particle, i.e. vectorial)
Poincar{\'e}, which includes Lorentz plus $P_\mu$ generators,
($M_{\mu\nu}; P_{\mu}$). For that case we show in Sect.2 that 1/2
BPS massive membrane representation contains all representations
of particle Poincar{\'e}, with given mass and different spins,
each spin appearing once. For 3/4 BPS (preon - \cite{Band})
representation, characterized by $k_{\alpha\beta}=\lambda_\alpha
\lambda_\beta$,  the answer is similar - after decomposition of
simplest (scalar) irrep of little group of preons we obtain all
massless representations of particle Poincar{\'e}, one for each
helicity. This last result can be interpreted as a group theory
point of view on Vasiliev's result \cite{Vas}. Although the
context is different - the $OSp(8)$ invariant equations of motion
in generalized space-time are considered, and problem of
construction of Cauchy surface is discussed in \cite{Vas},
mathematically the considerations are similar. The little group of
preons is $T_2$ - two-dimensional Euclidean translations. That
coincides exactly with the $T_2$ factor of little group of
massless particle with momenta of preon,
$p(k_{\alpha\beta})=p(\lambda_\alpha \lambda_\beta)$. Remind that
little group of massless particles is semidirect product of
$SO(2)$ with $T_2$, $SO(2)\ltimes T_2$.  So one can say that
excited states of preons correspond to the non-trivial
representations of that, usually trivially represented, factor of
little group of massless particles. Moreover, we can consider the
other subalgebra, namely $M_{\mu\nu},Z_{\mu\nu}$, i.e. that of
Lorentz generators plus membrane charge only. Mathematically it is
perfectly possible, but physically considerations of such an
algebras has to be justified. Particularly, it seems to be
impossible to write down the supersymmetry algebra with such
subalgebras, because the requirement of positivity of eigenvalues
of corresponding term in (\ref{4d}) can't be satisfied. This
differs from 12d susy algebra \cite{Bars} by signature, the two
time dimension in 12d make it possible to have susy algebra
without vector charge. But even in usual one-time signature case,
consideration of bosonic algebras is not forbidden by any general
considerations, and, particularly, the unitary irreps of that
algebras can be constructed. In that case we obtain the result,
that the little group of tensor
$z_{\mu\nu}(k_{\alpha\beta})=z_{\mu\nu}(\lambda_\alpha
\lambda_\beta)$ taken for preon representation  is same $T_2$
subgroup of 2d Euclidean translations factor of massless
particle's little group. So, excited (i.e. with non-trivial
representations of little group) representations of "pure
membrane" algebra ($M_{\mu\nu}; Z_{\mu\nu}$) corresponds to
excited $T_2$ generators of massless particles' little group in
usual Poincar{\'e}-invariant theories. Usually that generators are
represented trivially, by zero operators, because unitary irreps
of little group are required to be finite-dimensional which leads
to non-trivial representation of SO(2) subgroup of $SO(2)\ltimes
T2$ little group only, otherwise the unitary representation of
this noncompact group would be infinite dimensional. This sheds
some light on that factor of little group of massless particles
and one can conjecture, that corresponding excitations will appear
in full theory of 4d super Poincar{\'e} with tensorial charges.
One can conjecture, also, that self-consistent "pure branes"
theory may exist. Similar phenomena happens in higher dimensions.
In Sections 2, 3, 4, 5 we consider the decompositions of
representations of minimal tensorial Poincar{\'e} in dimensions
4,6,10 and 11 for different BPS states, mainly for preons, i.e.
those with maximal number of supersymmetries survived. In 11d case
(Section 5) a new phenomenon appears. In dimensions $d<11$, preons
are orbits of corresponding Lorentz group, i.e. are quotients G/H,
where H is the little group, a subgroup of Lorentz group G. So,
each spinor $\lambda_\alpha$ can be transformed into any other
spinor $\sigma_\alpha$, there is no different kinds of preons, and
one can speak about a preon representation of a
(super)Poincar{\'e}. At 11d space of spinors is 32 dimensional,
but orbits are generically 25 dimensional, so there are at least 7
invariants, distinguishing preons in 11d. Correspondingly, the
irreps of 11d (super) algebra will be labelled by that invariants.

In Conclusion results and prospects are discussed.

\section{(1+3)d }

Let's consider the minimal supersymmetry algebra in 4d Minkowski
space-time (\ref{4d}). That includes one Majorana spinor $Q$ (4
real components), Lorentz generators $M_{\mu\nu}$ (6),
energy-momentum vector $P_{\mu}$ (4) and brane (domain wall)
charge $Z_{\mu\nu}$ (6). Lhs of (\ref{4d}) is 4x4 real (more
exactly, with reality conditions, following from those on Majorana
spinor $Q$ in a given representation of gamma matrices, see below)
matrix with 10 independent components, which number coincides with
the number of independent real generators in rhs.

For construction of unitary irreps of corresponding bosonic
algebra $(M_{\mu\nu}; P_{\mu}, Z_{\mu\nu})$ we have to consider
orbits of Lorentz group in the space of vector and tensor,
construct a unitary irreps of stabilizer (little group) $H$ of a
given point on that orbit, and induce in a standard way that
representation to the representation of the whole Poincar{\'e} in
the space of functions on the orbit with values in a given irrep
of $H$. For an algebra (\ref{4d}) we shall consider the following
orbits: particle, when $Z_{\mu\nu}=0$; preon, when
$k_{\alpha\beta}=\lambda_{\alpha}\lambda_{\beta}$ ; BPS massive
membrane; "pure branes" orbit, with $P_\mu=0$ and different
$Z_{\mu\nu}$. Little groups for particle case are well known, some
other cases were considered in \cite{mrl2}, "pure branes" will be
calculated here for a first time, as well as different
decompositions of tensorial Poincar{\'e} irreps w.r.t. its
different bosonic subalgebras.

Consider first the particle case. It means that we are considering
usual Poincar{\'e} with $M_{\mu\nu}$ and $P_\mu$ generators. We
should consider orbits of Lorentz group on the space of momenta
$p$. Different orbits differs by values of Lorentz invariants,
$p^2$ is one of them. The physical cases are $p^2 \geq 0 $. For
$p^2=m^2>0$ little group is $SO(3)$, representations are
characterized by spin, integer or half-integer. For massless case
little group is two-dimensional Euclidean Poincar{\'e}, i.e. a
semidirect product of rotations $SO(2)$ and translations $T_2$ of
two-dimensional Euclidean plane, $SO(2)\ltimes T_2$. The known
interacting field theories are using the finite-dimensional
representations of $SO(2)\ltimes T_2$, representing translations
trivially, and representation of $SO(2)$ are classified by their
(integer or half-integer) helicity. As we shall see below,
non-trivial representations of translation generators corresponds
to states of "pure branes" theory.

Now consider the preon's state, i.e. states with
$k_{\alpha\beta}=\lambda_{\alpha}\lambda_{\beta}$, where
$\lambda_\alpha$ is commuting Majorana 4d spinor. One calculation
gives us simultaneously the statement that space of
$\lambda_\alpha$ is a homogeneous space of Lorentz group $SO(1,3)$
and an algebra of its little group. Namely, we act by Lorentz
generators on $\lambda_\alpha $ and find that its stabilizer is
$T_2$ group, and dimensionality of orbit is 4, i.e. equal to that
of whole space of spinors. We use Weyl representation of
gamma-matrices:
\begin{eqnarray}
\Gamma ^{\mu}=\begin{pmatrix}
  0 & \sigma_{\mu} \\
  \bar{\sigma}_{\mu} & 0
\end{pmatrix}\\
\sigma_{\mu}=(1,\sigma_i),\bar{\sigma}_{\mu}=(-1,\sigma_i)
\end{eqnarray}
where $\sigma^i$ are Pauli matrices. The similar relation defines
our gamma matrices in any even dimension through gamma matrices of
previous Euclidean dimension. Then (pseudo)Majorana condition on
spinor can be  deduced (see, e.g. \cite{kugo}), 4d Majorana spinor
$\lambda_{\alpha}$ satisfies
\begin{eqnarray}
(\lambda_{\alpha})^*=\Gamma ^{0}\Gamma ^{1}\Gamma
^{3}\lambda_{\beta}
\end{eqnarray}

Then stabilizer algebra of e.g. Majorana spinor (1,0,0,1) is

\begin{eqnarray}
\begin{pmatrix}
  0 & a &b&0 \\
  -a & 0&0&a\\
  -b&0&0&b\\
  0&-a&-b&0 \label{T2}
\end{pmatrix}
\end{eqnarray}

This corresponds to non-compact group $T_2$ of translations which
coincides with that of massless particle, because little group of
particle with $p^2= p(k)^2$ should contain that for $k$ as
subgroup. This is right for any $k$, because stabilizer of $k$ is
intersection of stabilizers of $p_\mu(k)$ and $z_{\mu\nu}(k)$. In
our case $p(k)^2=0$, so little group for massless particle
$SO(2)\ltimes T_2$ contains as subgroup that of preons, i.e.
$T_2$. These considerations are confirmed by dimensionality check:
the orbit in four-momenta space is 3-dimensional cone
$SO(1,3)/(SO(2)\ltimes T_2)$, and preon orbit is four dimensional
$SO(1,3)/T_2$. So, the particle representation is induction to the
whole Poincar{\'e} of the representation of little group in the
space of functions on $SO(1,3)/(SO(2)\ltimes T_2)$ with values in
unitary irreps of $SO(2)\ltimes T_2$, irreps of preons are
induction to the whole tensorial Poincar{\'e} from the
representation of the little group in the space of functions on
$SO(1,3)/T_2$, with values in unitary irreps of $T_2$. This last
orbit is a fiber bundle over the first one with SO(2) fiber. For
decomposition of representation of tensorial $(M_{\mu\nu};
P_{\mu}, Z_{\mu\nu})$ Poincar{\'e} w.r.t. the usual $(M_{\mu\nu};
P_{\mu})$ Poincar{\'e} we have to decompose the space of functions
on that bundle into the space of functions on the base. So, we
have to find the space of spinors, giving the same momenta
$p_\mu$, i.e. fiber over given $p_\mu$, and define the action of
$SO(2)$ on that fiber. Decomposition of the space of function on
that fiber w.r.t. $SO(2)$ gives the helicity content of preon
representation. The space of spinors giving same $p_\mu$, say
$p_\mu=(1,0,0,1)$ is $\lambda_2=\lambda_3=0, \lambda_4=
\lambda_1^*,  |\lambda_1|=1 $. $SO(2)$ transformations are acting
on $\lambda_1$ by phase rotations, and space of functions gives
the whole integer spectrum of helicities, each one once. For the
space of double-valued functions (spinor type functions) the
spectrum will consist of all half-integer helicities, each
helicity appearing once. This result actually is identical to that
of Vasiliev \cite{Vas}, as mentioned above.

It is natural to consider another subgroup of 4d tensorial
Poincar{\'e}, namely $(M_{\mu\nu}; Z_{\mu\nu})$. In that case we
have to calculate $Z_{\mu\nu}$ for preons, find its little group
and decompose preon's little group representation w.r.t. that
little group, i.e. decompose space of functions on a fiber over
given $z_{\mu\nu}(k)$ w.r.t. that little group. Direct calculation
for given $k_{\alpha}\!^{\beta}$ gives
\begin{eqnarray}
z_{\mu\nu}(k)=
\begin{pmatrix}
  0 & -1 &0 &0 \\
 1 & 0 &0 & 1 \\
  0& 0 & 0 & 0 \\
  0 & -1 & 0 & 0
\end{pmatrix}
\end{eqnarray}
The stabilizer of this matrix is the same $T_2$ (\ref{T2}). So
actually there is no fiber over this point, and irrep of
$(M_{\mu\nu};P_\mu,Z_{\mu\nu})$ gives an irrep of
$(M_{\mu\nu};Z_{\mu\nu})$.

The above analysis can be repeated for other, not preon
representations of tensorial Poincar{\'e}
$(M_{\mu\nu};P_\mu,Z_{\mu\nu})$. Take a BPS membrane state, with
\begin{equation}
    P_\mu=(m,0,0,0),  Z_{12}=-Z_{21}=m  \label{mmb}
\end{equation}
This is a 1/2 massive (membrane) BPS of susy algebra (\ref{4d}).
One can find a little group for that configuration (\cite{mrl}),
it is SO(2) (rotations around 3-rd axis), so orbit is
SO(1,3)/SO(2). The little group for particle subalgebra is
$SO(3)$, with orbit $SO(1,3)/SO(3)$. Correspondingly, fiber is
$SO(3)/SO(2)=S^2$. So, for an e.g. representation, induced from
trivial representation of SO(2) we have to decompose the space of
functions on $S^2$ w.r.t. its invariance group SO(3). That is the
sum of all representations of SO(3), with any spin, with
multiplicity one. Decomposition w.r.t. the "pure membrane"
subalgebra $(M_{\mu\nu},Z_{\mu\nu})$ can be done provided we
define the little group for tensor $z_{\mu\nu}$. That is
$SO(2)\otimes T_1$, so orbit is $SO(1,3)/SO(2)\otimes T_1$, and
fiber for decomposition of membrane representation on a "pure
membrane" representations is $T_1$.

\section{6d}

The 6d Minkowski space supersymmetry algebra we shall consider
includes, besides Lorentz generators $M_{\mu\nu}$, the Weyl spinor
$Q_{\alpha}$ (8 real components), vector $P_{\mu\nu}$ (6) and
third rank self-dual tensor $Z^+_{\mu\nu\lambda}$ (10). Susy
relation is
\begin{eqnarray}
\left\{ \bar{Q},Q\right\} &=&\Gamma ^{\mu}P_{\mu}+\Gamma
^{\mu\nu\lambda}Z^+_{\mu\nu\lambda},  \label{6d}\\
  \mu,\nu,...
&=&0,1,2,3,4,5. \nonumber
\end{eqnarray}
The lhs is 4x4 Hermitian matrix, with 16 real components, which
coincides with number of real generators in rhs. Usually in 6d one
considers symplectic Majorana-Weyl spinors, which, while having
same number of real components, have a bigger invariance group -
SU(2) instead of U(1) in our Weyl spinors case. But last one is
simpler, and we consider that here.

The preon representation of (\ref{6d}) corresponds to r.h.s.
matrix of rank one. That can be parameterized as $k=\lambda
\bar{\lambda}$, where Weyl spinor $\lambda$ is defined up to phase
transformation. The space of Weyl spinors $\lambda$ is a
homogeneous space $SO(1,5)/SO_L(3)\ltimes T_4$, where
$SO(3)_L\ltimes T_4$ is the stabilizer of spinors. The little
group of preon representation, i.e. stabilizer of r.h.s. of
(\ref{6d}), $\lambda \bar{\lambda}$, has an additional SO(2)
factor: $(SO_L(3)\times SO_R(2)\ltimes T_4)$. Namely, let's take a
Weyl spinor of e.g. form $(1,0,0,0)$, and transform that under
SO(1,5) rotation. Then in our representation of gamma matrices the
algebra of stabilizer of this spinor is given by following matrix:

\begin{eqnarray}
\begin{pmatrix}
  0 & -w_{2} & -w_{3} & -w_{4} & -w_{5} & 0 \\
  w_{2} & 0 & w_{23} & w_{24} & w_{25} & w_{2} \\
  w_{3} & -w_{23} &0 & w_{25} & -w_{24} & w_{3} \\
  w_{4} & -w_{24} & -w_{25} & 0 & w_{23} & w_{4} \\
  w_{5} & -w_{25} & w_{24} & -w_{23} & 0 & w_{5} \\
  0 & -w_{2} & -w_{3} & -w_{4} & -w_{5} & 0 \label{6d}
\end{pmatrix}
\end{eqnarray}
which is a semidirect product of Abelian algebra of translations
$T_4$ (with parameters $w_{2}, w_{3}, w_{4}, w_{5}$) and algebra
$so_R(3)$ (remaining $w$-s). Algebra of phase transformations of a
given spinor $(1,0,0,0)$ in our representation of gamma-matrixes
are given by $so_L(2)$.

The little group of 6d massless particles is $SO(4)\ltimes T_4$
(given by  (\ref{6d}) without restrictions on the elements of
middle 4 by for 4 antisymmetric matrix) so fiber for decomposition
of preon representation w.r.t. the particles representations is
$SO(4)/(SO_R(3)\times SO_L(2) \backsim SO(3)/SO(2)$. So, we have
to decompose the space of functions (considering simplest
representation, with little group represented trivially) on
$SO(4)/(SO_R(3)\times SO_L(2))\backsim SO(3)/SO(2)$ w.r.t. the
SO(4) rotations. As we see, one of SO(3) factors of SO(4) (we
consider the covering group) is represented trivially, for other
SO(3) factor representations of all spins appear, with
multiplicity one.

Similarly to previous Section, we can consider "pure 3-brane"
algebra, i.e. $(M_{\mu\nu}, Z_{\mu\nu\lambda})$ algebra. In that
case for our spinor $(1,0,0,0)$ the $Z_{\mu\nu\lambda}$ tensor has
non-zero components $Z_{125}=-Z_{134}=Z_{256}=-Z_{346}$ (and those
with permutations of indexes) only, and its little group is
($SO_L(3)\times SO_R(2))\ltimes T_4$, i.e. coincide with that of
preons. So the fiber for decomposition of preon representation
w.r.t. 3-brane algebra is trivial, and one has one "pure 3-brane"
irrep in decomposition of preon irreps.

\section{10d}

For 10d Minkowski space minimal supersymmetry algebra includes,
besides Lorentz generators $M_{\mu\nu}$,  Majorana-Weyl spinor $Q$
(16 real components), vector $P_{\mu}$ (10) and fifth rank
self-dual tensor $Z^+_{\mu\nu\lambda\rho\sigma}$ (126). Susy
relation is given by

\begin{eqnarray}
\left\{ \bar{Q},Q\right\} &=&\Gamma ^{\mu}P_{\mu}+\Gamma
^{\mu\nu\lambda\rho\sigma}Z^+_{\mu\nu\lambda\rho\sigma},  \label{10d}\\
  \mu,\nu,...
&=&0,1,2,...,9 \nonumber
\end{eqnarray}

The lhs is 16x16 symmetric (after right multiplication on charge
conjugation C matrix) matrix, with 136 real components, which
coincide with count of real generators in rhs. Consider
corresponding bosonic algebra
$M_{\mu\nu};P_{\mu},Z^+_{\mu\nu\lambda\rho\sigma}$ and its preon
representation, when rhs (after C multiplication) is
$\lambda_\alpha\lambda_\beta$. The 16d space of spinors
$\lambda_\alpha$ is homogeneous space $SO(1,9)/SO(7)\ltimes T_8$,
where action of $SO(7)$ on $T_8$ is defined by its insertion, in
its spinorial representation, into adjoint representation of
$SO(8)$. So, the little group is $SO(7)\ltimes T_8$. In our gamma
matrices representation the above statements appear as follows:
the Weyl condition is selecting first 16 components of general 32d
spinor, the Majorana condition requires $Q_1=Q_6^*, Q_2=-Q_5^*,
Q_3=-Q_8^*, Q_4=Q_7^*$. Then calculation of stabilizer of specific
spinor with, e.g. non-zero and equal to 1 first and sixth
components only gives a Lorentz generator matrix
\begin{eqnarray}
\begin{pmatrix}
  0 & -w_{2} & -w_{3} & -w_{4} & -w_{5} & -w_{6} & -w_{7} & -w_{8} & -w_{9} & 0 \\
  w_{2} & 0 & -c_{1} & -c_{2} & -c_{3} & -c_{4} & -c_{5} & -c_{6} & -c_{7} & w_{2} \\
  w_{3} & c_{1} & 0 & w_{34} & w_{35} & w_{36} & w_{37} & w_{38} & w_{39} & w_{3} \\
  w_{4} & c_{2} & -w_{34} & 0 & w_{45} & w_{46} & w_{47} & w_{48} & w_{49} & w_{4} \\
  w_{5} & c_{3} & -w_{35} & -w_{45} & 0 & w_{56} & w_{57} & w_{58} & w_{59} & w_{5} \\
  w_{6} & c_{4} & -w_{36} & -w_{46} & -w_{56} & 0 & w_{67} & w_{68} & w_{69} & w_{6} \\
  w_{7} & c_{5} & -w_{37} & -w_{47} & -w_{57} & -w_{67} & 0 & w_{78} & w_{79} & w_{7} \\
  w_{8} & c_{6} & -w_{38} & -w_{48} & -w_{58} & -w_{68} & -w_{78} & 0 & w_{89} & w_{8} \\
  w_{9} & c_{7} & -w_{39} & -w_{49} & -w_{59} & -w_{69} & -w_{79} & -w_{89} & 0 & w_{9} \\
  0 & -w_{2} & -w_{3} & -w_{4} & -w_{5} & -w_{106} & -w_{7} & -w_{8} & -w_{9} &
  0 \label{10L}
\end{pmatrix}
\end{eqnarray}
where $c_1=-w_{46} - w_{57} + w_{89}, c_2=w_{36} + w_{58} +
w_{79}, c_3=w_{37} - w_{48} - w_{69}, c_4=-w_{34} + w_{59} -
w_{78}, c_5=-w_{35} - w_{49} + w_{68}, c_6=-w_{39} + w_{45} -
w_{67}, c_7=w_{38} + w_{47} - w_{56} $.

The structure of algebra (\ref{10L}) is the following: it is the
semidirect product of $T_8$ - Abelian algebra of matrices
(\ref{10L}) with non-zero first row, last row, first and last
columns only, and remaining matrices. This last algebra,
represented by middle 8x8 matrices of (\ref{10L}, is spinor
representation of SO(7), based on a real representation of 7d
gamma matrices. Such representation can be constructed with the
help of multiplication table of octonions, as shown in \cite{Oct}.

The little group for massless particle is $SO(8)\ltimes T_8$, the
fiber for decomposition of preon representation of is
$SO(8)/SO(7)=S^7$. Correspondingly, the decomposition of simplest
representation of preons, i.e. that in the space of scalar
functions on orbit, with respect to particle subalgebra
$(M_{\mu\nu};P_{\mu})$ is given by the sum of symmetric tensor
representations with multiplicity 1 (\cite{BR}, Section 10.3).

\section{11d}

For 11d Minkowski space supersymmetry algebra includes Lorentz
generators $M_{\mu\nu}$, the Majorana spinor $Q_\alpha$ (32 real
components), vector $P_\mu$ (11), second rank tensor $Z_{\mu\nu}$
(55) and fifth rank tensor $Z_{\mu\nu\lambda\rho\sigma}$ (462).
Anticommutator of supercharges is given by (\ref{eq1}). The lhs is
32x32 real symmetric matrix, with 528 real components, which
coincide with count of real generators in rhs. The strong
difference with previous cases is in that  space of 11d spinors is
not a homogeneous manifold of Lorentz group $SO(1,10)$, so it is
not correct to define the representation by simply stating that
rank of rhs of (\ref{eq1}) is one, hence it is equal to
$\lambda_{\alpha}\lambda_{\beta}$ with some spinor
$\lambda_{\alpha}$. One has to define the values of additional
invariants which separate the homogeneous sub-manifolds in the
space of spinors $\lambda_{\alpha}$. The number of such invariants
should be at least 7, because stabilizer of e.g. spinor with
non-zero (and equal to 1) entries at first and sixth places only
(this is a particular Majorana spinor in our gamma matrices
representation, see previous Section) is 30-dimensional group
$SO(7)\ltimes T_9$, so dimensionality of quotient
$SO(1,10)/SO(7)\ltimes T_9$ is $55-30=25$, which is 7 units less
than dimensionality of spinor's space. So, one has at least 7
kinds of preons, the values of invariants, which distinguish them,
simultaneously are labelling irreps of tensorial Poincar{\'e}
$(M_{\mu\nu};P_\mu,Z_{\mu\nu},Z_{\mu\nu\lambda\rho\sigma})$. For
the orbit considered we can look for a stabilizer of vector
$P_\mu$ and second-rank tensor $Z_{\mu\nu}$, i.e. consider the
decomposition w.r.t. the $(M_{\mu\nu};P_\mu,Z_{\mu\nu})$
subalgebra. The corresponding stabilizer is $SO(8)\ltimes T_9$, so
fiber is $SO(8)/SO(7)=S^7$, and according to general results
(\cite{BR}, Section 10.3), for the simplest case of trivial
representation of little group, the space of functions on $S^7$
contains all symmetric tensors representations (one row Young
diagram), each one once. Next, it is not difficult to define
stabilizer of second rank membrane tensor $Z_{\mu\nu}$, which is
necessary for decomposition w.r.t. the "pure membrane" subalgebra
$(M_{\mu\nu},Z_{\mu\nu})$. That is again $SO(8)\ltimes T_9$.
Finally, we comment on the decomposition of preon's state w.r.t.
the massless particle (which corresponds to preons). Since little
group for massless particle is $SO(9)\ltimes T_9$ for
decomposition of preons w.r.t. particles we obtain a fiber
$SO(9)/SO(7)$. The final answer can be obtained both directly, by
decomposing the space of functions on $SO(9)/SO(7)$ w.r.t. SO(9)
group, or, using previous result, in two stages - first
decomposing w.r.t. particle + membrane subalgebra, i.e. taking
fiber SO(8)/SO(7) and then decomposing the space of functions on
fiber SO(9)/SO(8), with values in each of representations,
obtained on a previous stage. According to the general theorem
\cite{BR}, results should be the same.

\section{Conclusion}
We have calculated little groups (algebras) for different orbits
of tensorial Poincar{\'e} algebras at different dimensions. That
groups are useful in construction of irreps of corresponding
(super)Poincar{\'e} algebras with tensorial charges, in discussion
of spin-statistics for branes \cite{mrl2}, etc. We use these
results for decomposition of scalar irreps of tensorial
Poincar{\'e} algebra w.r.t. the proper subalgebras with highest
rank tensor removed. I.e. for d=4,6,10 it is the decomposition
w.r.t. the subalgebra $M_{\mu\nu};P_{\mu}$, at 11d complete result
is obtained for subalgebra $M_{\mu\nu};P_{\mu},Z_{\mu\nu}$.
Results obtained permit the consideration of other cases, also. It
seems that present approach provide group's theory point of view
on Vasiliev's results \cite{Vas}, and can be helpful there in
higher dimensions. We show, that at d=11 preon \cite{Band} states
are not defined by simply stating that r.h.s of (\ref{eq1}) has
rank one, and is square of some spinor, but one should define the
specific orbit to which that spinor belongs, which generally
requires definition of 7 parameters. This fact requires further
study of what is real difference in representations of tensorial
Poincar{\'e} for different preon orbits. Also results on a little
groups can be extended to other subalgebras of tensorial
Poincar{\'e} algebras, as well as to other algebras, corresponding
to theories with extended supersymmetries, or (6d case)
formulations of the same algebra with different symmetry group.
The most relevant next steps are extension of construction
\cite{Man1}, \cite{Man2}, \cite{mrl} of field theories with
tensorial Poincar{\'e} algebra as space-time symmetry on some of
representations, discussed in the present paper.

\section{Acknowledgements}

This work is supported partially by INTAS grant \#99-1-590. I'm
indebted to R.Manvelyan for discussions.


\begin{thebibliography}{9}
\bibitem{TowAsk} J.A.de Azcarraga, J.P.Gauntlett, J.M.Izquirdo and
P.K.Townsend, Phys.Rev.Lett. 63 (1989) 2443
\bibitem{Tow} P.K. Townsend, $M$-theory from its superalgebra,
hep-th/9712004.
\bibitem{Wig} E. Wigner, Ann. Math. 40, 149 (1939)
\bibitem{BR} A.O.Barut and R.Raczka, Theory of Group
Representations and Applications, PWN Warszawa, 1977,
\bibitem{Man1}  R. Manvelyan and R. Mkrtchyan, Mod.Phys.Lett. A15 (2000)
747-760, hep-th/9907011.
\bibitem{Man2}  R. Manvelyan and R. Mkrtchyan, Mod.
Phys. Lett. A, Vol. 17, No. 21 (2002) pp.1393-1406,
hep-th/0112233.
\bibitem{Bars} I. Bars,  Phys.Rev. D54 (1996)
5203-5210, hep-th/9604139 \newline I. Bars, Duality and hidden
dimensions, hep-th/9604200, \newline I.Bars, Phys.Rev. D55 (1997)
2373-2381, hep-th/9607112
\bibitem{mrl2} R.Mkrtchyan, Little Groups and Statistics of Branes,
hep-th/0209040.
\bibitem{mrl} R.Mkrtchyan, On an (interacting) field theories with
tensorial momentum, Report on 3-rd Sakharov Conference (Moscow,
June 24-29, 2002), hep-th/0209175.
\bibitem{Vas} M.A.Vasiliev, Phys.Rev. D66 (2002) 066006,
hep-th/0106149.
\bibitem{Vas2} M.A.Vasiliev, Relativity, Causality, Locality,
Quantization and Duality in the Sp(2M) Invariant Generalized
Space-Time, Contribution to the Marinov's Memorial Volume,
M.Olshanetsky and A.Vainshtein, Eds, World Scientific,
hep-th/0111119,

\bibitem{Band}Igor A. Bandos, Jos'e A. de Azc'arraga, Jos'e M.
Izquierdo and Jerzy Lukierski, Phys.Rev.Lett. 86 (2001) 4451-4454,
hep-th/0101113.
\bibitem{Band2} Igor Bandos  and Jerzy Lukierski, Mod.Phys.Lett.
A14 (1999) 1257-1272, hep-th/9811022,
\newline I.Bandos, J.Lukierski and D.Sorokin, Phys. Rev.D61
(2000) 45002, hep-th/9904109.  \newline Jerome P. Gauntlett and
Chris M. Hull,  JHEP 0001 (2000) 004, hep-th/9909098, \newline
Jerome P. Gauntlett, Gary W. Gibbons, Christopher M. Hull and Paul
K. Townsend, Commun.Math.Phys. 216 (2001) 431-459, hep-th/0001024.

\bibitem{kugo} T.Kugo and P.Townsend, Nucl.Phys.B221, (1983) 357
\bibitem{Oct} Murat Gunaydin and Hermann Nicolai, Phys.Lett. B351
(1995) 169-172, Addendum-ibid. B376 (1996) 329, hep-th/9502009
\newline Murat Gunaydin and Sergei V. Ketov, Nucl.Phys. B467
(1996) 215-246, hep-th/9601072 \newline J.C. Baez, The Octonions,
math.RA/0105155
\newline Khaled Abdel-Khalek, Ring Division Algebras, Self-Duality
and Supersymmetry, hep-th/0002155

\end{thebibliography}
\end{document}